\documentclass[twocolumn,prc,aps,floatfix]{revtex4}
\usepackage[dvips]{graphicx}
\begin{document}
\title{Limits on temporal variation of fine structure constant,
 quark masses and strong interaction from atomic clock experiments}
\author{V.V. Flambaum}
\affiliation{
 School of Physics, The University of New South Wales, Sydney NSW
2052, Australia
}
\date{\today}
\begin{abstract}
We perform calculations of the dependence of nuclear magnetic moments
on quark masses and
obtain limits on the variation of
$(m_q/\Lambda_{QCD})$
 from recent atomic clock
experiments with hyperfine transitions in H, Rb, Cs, Yb$^+$, Hg$^+$
and optical transition in Hg$^+$. Experiments  with
Cd$^+$,  deuterium/hydrogen, molecule SF$_6$,
 Zeeman transitions in $^3$He/Xe are also discussed.
\end{abstract}
\maketitle


\section{Introduction}
 Interest in the temporal and spatial variation of  major constants of physics
has been recently revived by
 astronomical data which seem to suggest a variation
 of the electromagnetic constant
 $\alpha=e^2/\hbar c$   at the $10^{-5}$ level
 for the time scale 10 billion years, see \cite{alpha}
 (a discussion of
other limits can be found in the review \cite{uzan} and references therein).
 However, an independent experimental confirmation is needed.

The hypothetical unification of all interactions implies that variation
of the electromagnetic interaction constant $\alpha$ should be accompanied
by the variation of masses and the strong interaction constant.
 Specific predictions need a model. For example, the grand unification
model  discussed in \cite{Langacker:2001td} predicts that
 the
quantum chromodynamic (QCD) scale  $\Lambda_{QCD}$
 (defined as the position of the Landau pole in the logarithm for the
running strong coupling constant)
is modified as follows: $\delta \Lambda_{QCD} / \Lambda_{QCD}
\approx 34 \delta \alpha / \alpha$.
The variation of quark and electron masses in this model is  given by
$\delta m / m \sim 70 \delta \alpha / \alpha $.
This gives an estimate for the variation of the dimensionless ratio
\begin{equation} \label{mQCD}
{\delta(m/ \Lambda_{QCD}) \over(m/\Lambda_{QCD})}\sim 35 {\delta \alpha
\over \alpha}
\end{equation}
This result is strongly model-dependent.
However, the large coefficients in these expressions are generic for
 grand unification models, in which modifications come from high energy scales:
they appear because the running strong coupling constant and
 Higgs constants (related to mass) run faster than $\alpha$.
 This means that if these models
are correct the variation
of masses and strong interaction may be easier to detect than the variation
of $\alpha$.

One can measure only variation of the
dimensionless quantities, therefore we want to extract from the measurements
 variation
of the dimensionless ratio $m_q/\Lambda_{QCD}$ where $m_q$ is the quark
mass (with the dependence on the normalization point removed).
A number of limits on variation of $m_q/\Lambda_{QCD}$
have been obtained recently from consideration
of Big Bang Nucleosynthesis, quasar absorption spectra
and Oklo natural nuclear reactor which was active about
1.8 billion years ago \cite{FS,oliv,dmitriev,FS1} (see also
\cite{Murphy1,Cowie,Oklo,c12,savage}). Below we consider
the limits which follow from laboratory atomic clock comparison.
Laboratory limits with a time base about a year are
especially sensitive to oscillatory variation of
fundamental constants. A  number of relevant measurements
have been performed already and even larger number  have been started
or planned. The increase in precision is very fast. 

It has been pointed out by Karshenboim \cite{Karschenboim}
 that measurements  of ratio of hyperfine structure intervals
in different atoms are sensitive to
variation of nuclear magnetic moments.
 First rough estimates of the dependence of nuclear magnetic
 moments  on 
 $m_q/\Lambda_{QCD}$ and limits on time variation of this ratio
have been obtained in our paper
\cite{FS}. Using 
H, Cs and Hg$^+$ measurements \cite{prestage,Cs},
 we obtained the limit on variation of $m_q/\Lambda_{QCD}$
 about $5 \cdot 10^{-13}$ per year.
Below we calculate the dependence of nuclear magnetic moments
 on  $m_q/\Lambda_{QCD}$ and  obtain the limits 
 from recent atomic clock
experiments with hyperfine transitions in H, Rb, Cs,Yb$^+$,Hg$^+$
and optical transition in Hg$^+$.
It is convenient to assume that the strong interaction scale
 $\Lambda_{QCD}$ does not vary, so we will speak about variation
of masses. We shall restore $\Lambda_{QCD}$ in final answers.
   
     The hyperfine structure constant can be presented in the following
form
\begin{equation}\label{A}
A=const \times [\frac{m_e e^4}{\hbar ^2}] [ \alpha ^2 F_{rel}( Z \alpha)]
[\mu \frac{m_e}{m_p}]
\end{equation}
The factor in the first bracket is an atomic unit of energy. The second 
``electromagnetic'' bracket determines the dependence on $\alpha$.
An approximate expression for the relativistic correction factor (Casimir
factor) for s-wave electron is the following
\begin{equation}\label{F}
F_{rel}= \frac{3}{\gamma (4 \gamma^2 -1)}
\end{equation}
where $\gamma=\sqrt{1-(Z \alpha)^2}$, Z is the nuclear charge.
Variation of $\alpha$ leads to the following variation of $F_{rel}$
 \cite{prestage}:
\begin{equation}\label{dF}
\frac{\delta F_{rel}}{F_{rel}}=K \frac{\delta \alpha}{\alpha}
\end{equation}
\begin{equation}\label{K}
K=\frac{(Z \alpha)^2 (12 \gamma^2 -1)}{\gamma^2 (4 \gamma^2 -1)}
\end{equation}
More accurate numerical many-body calculations \cite{dzuba1999}
 of the dependence of the hyperfine structure on $\alpha$ have shown
 that the coefficient $K$ is slightly larger than that given by this
formula. For Cs ($Z$=55) $K$= 0.83 (instead of 0.74),
for Rb $K$=0.34
(instead of 0.29),  for Hg$^+$
$K$=2.28 (instead of 2.18).

    The last bracket in eq. (\ref{A})  contains the dimensionless
 nuclear magnetic moment $\mu$ in nuclear magnetons
 ( the nuclear magnetic moment $M=\mu\frac{e\hbar}{2 m_p c}$),
 electron mass $m_e$ and proton mass $m_p$. We may also include
a small correction due to the finite nuclear
size. However, its contribution is insignificant.

Recent experiments measured time dependence of the ratios of
 hyperfine structure intervals of $^{199}$Hg$^+$ and H \cite{prestage},
$^{133}$Cs and $^{87}$Rb \cite{marion} and ratio of optical frequency
 in Hg$^+$ and $^{133}$Cs hyperfine frequency \cite{bize}.  
 In the ratio of two
hyperfine structure constants for different atoms time dependence
may appear from the ratio of the factors $F_{rel}$ (depending on $\alpha$)
 and ratio of nuclear magnetic moments (depending on $m_q/\Lambda_{QCD}$).
Magnetic moments in a single-particle approximation (one unpaired nucleon)
 are:  
\begin{equation}\label{mu+}
\mu=(g_s + (2 j-1) g_l)/2
\end{equation}
for $j=l+1/2$.
\begin{equation}\label{mu-}
\mu=\frac{j}{2(j+1)}(-g_s + (2 j+3) g_l)
\end{equation}
for $j=l-1/2$. Here the orbital g-factors are
 $g_l=1$ for valence proton and $g_l=0$ for valence
neutron. The present values of spin g-factors $g_s$ are
$g_p=5.586$ for proton and $g_n=-3.826$ for neutron.
 They depend on $m_q/\Lambda_{QCD}$.
The light quark masses are only about $1 \%$ of the nucleon mass
 ($m_q=(m_u+m_d)/2 \approx$ 5 MeV). The nucleon magnetic moment remains
 finite in the chiral  limit of $m_u=m_d=0$. Therefore, one may think that 
 the corrections to $g_s$ due to the finite quark masses are very small.
However, there is a mechanism which enhances quark mass contribution:
$\pi$-meson loop corrections to the nucleon magnetic moments which
are proportional to $\pi$-meson mass $m_{\pi} \sim \sqrt{m_q\Lambda_{QCD}} $;
$m_{\pi}$=140 MeV is not so small.

    According to  calculation in Ref. \cite{Thomas} dependence of the
nucleon g-factors
on $\pi$-meson mass $m_\pi$ can be approximated by the following equation
\begin{equation}\label{thomas}
g(m_\pi)=\frac{g(0)}{1+ a m_\pi + b m_\pi ^2}
\end{equation}
where $a$= 1.37/GeV ,  $b$= 0.452/GeV$^2$ for proton
and  $a$= 1.85/GeV ,  $b$= 0.271/GeV$^2$ for neutron.
This formula is a fit of the numerical calculations performed
using the Cloudy Bag Model (pion field coupled to nucleon quarks).
This model reproduces leading terms of chiral perturbation theory
for nucleon magnetic moments \cite{Savage1} (small $m_\pi$ limit).
The results also agree with lattice calculations \cite{Thomas} (performed at
 $m_\pi \sim 500$ MeV).  
Eq. (\ref{thomas}) leads to
the following estimates:
\begin{equation}\label{gp}
\frac{\delta g_p}{g_p} =
 -0.174 \frac{\delta m_\pi}{m_\pi}= -0.087 \frac{\delta m_q}{m_q}
\end{equation}
\begin{equation}\label{gn}
\frac{\delta g_n}{g_n} =
 -0.213 \frac{\delta m_\pi}{m_\pi}= -0.107 \frac{\delta m_q}{m_q}
\end{equation}  
Eqs. (\ref{mu+},\ref{mu-},\ref{gp},\ref{gn}) give variation
 of nuclear magnetic moments as functions of  $m_q/\Lambda_{QCD}$.

The measured time variations of ratios of hyperfine frequencies depend
 on two parameters: the ratio of proton spin  magnetic moment and 
proton orbital magnetic moment (this ratio is proportional to $g_p$) and 
the ratio of proton and neutron spin magnetic moments ($M_p/M_n= g_p/g_n$).
According to eqs. (\ref{gp},  \ref{gn}) the ratio
$ g_p/g_n$ practically does not depend on $m_q$
and seems to be not sensitive to variation of quark masses and strong interaction.
However, this conclusion may be misleading. Magnetic moments
depend also on the strange quark mass $m_s$. In the minimal
order of the chiral perturbation theory this dependence
is dominated by  mixing $p - K^+ \Lambda$. Similar process involving
$K^0$ meson loop and $\Lambda$ does not contribute to the  neutron
magnetic moment since $K^0$ does not carry electric charge. As a result
lowest order  corrections for proton and neutron magnetic moments
are very different: for proton $a(K)/a(\pi)=0.4$, for neutron
 $a(K)/a(\pi)=-0.03$ \cite{Savage1,Thomas1} (see eq. (\ref{thomas})
for  definition of the coefficient $a$). The proton (or neutron) g-factor 
is a ratio of the proton (or neutron) spin magnetic moment to
the  nuclear magneton $e\hbar/2 m_p c$
(quantum for proton orbital magnetic moment).
 Therefore,  dependence of $g_p \sim M_p m_p$
and $g_n \sim M_n m_p$ on the strange
quark mass may  also appear from the relatively large ($\sim 20 \%$)
contribution of $m_s$ to the proton mass $m_p$ \cite{FS,dmitriev}.
In the case of $g_p$ this contribution is of opposite sign to that 
of the lowest  order in the chiral perturbation theory ($a(K) m_K$) and has
 comparable
 magnitude. This cancellation means that we can not reliably determine
dependence of $g_p$ 
on the strange quark mass $m_s$. In this case it would be safer
to  neglect contribution of the strange quark mass to $g_p$.
However, we want to keep dependence
on the strange quark mass in the ratio of neutron and proton magnetic
moments since in a number of cases this is the only effect which
provides dependence on fundamental  masses and strong interaction.
Using  the lowest order chiral perturbation theory results
 \cite{Savage1,Thomas1} presented above we obtain the
 dependence on the strange quark
mass $\frac{\delta(g_n/ g_p)}{g_n/g_p} \approx 0.1\frac{\delta m_s}{m_s} $.
 This result should possibly be treated as an order of magnitude estimate
since  $K$- meson mass is not as small as $\pi$-meson mass and
 accuracy of the chiral  theory may be low in this case.
Thus, we arrive to the following equations:
\begin{equation}\label{gps}
\frac{\delta g_p}{g_p} = -0.087 \frac{\delta m_q}{m_q}
\end{equation}
\begin{equation}\label{gns}
\frac{\delta g_n}{g_n} = -0.107 (\frac{\delta m_q}{m_q}
 -\frac{\delta m_s}{m_s})
\end{equation}  
Note that transferring the entire $m_s$-dependence into
the neutron g-factor is mostly a matter of convenience,
all that we can say is that $g_n/g_p$ depends on $m_s/\Lambda_{QCD}$.  
Now we can find variation of nuclear magnetic moments using
 eqs. (\ref{mu+},\ref{mu-}). For all even Z nuclei with valence
neutron ($^{199}$Hg,$^{171}$Yb,$^{111}$Cd, etc)
 we obtain $\frac{\delta \mu}{\mu} =\frac{\delta g_n}{g_n}$.
For $^{133}$Cs we have valence proton with $j$=7/2, $l$=4 and 
\begin{equation}\label{Cs}
\frac{\delta \mu}{\mu} =
 0.22 \frac{\delta m_\pi}{m_\pi}= 0.11 \frac{\delta m_q}{m_q}
\end{equation}
For $^{87}$Rb we have valence proton with $j$=3/2, $l$=1 and 
\begin{equation}\label{Rb}
\frac{\delta \mu}{\mu} =
 -0.128 \frac{\delta m_\pi}{m_\pi}=- 0.064 \frac{\delta m_q}{m_q}
\end{equation}

 As an intermediate result it is convenient to present dependence of
 the ratio of hyperfine
constant $A$ to atomic unit of energy  $E=\frac{m_e e^4}{\hbar ^2}$ (or
 energy of 1s-2s transition in hydrogen equal to 3/8 $E$)
on variation of the fundamental constants. We introduce a parameter $V$
defined by the realation
\begin{equation}\label{V}
\frac{\delta V}{V} \equiv \frac{\delta (A/E)}{A/E}
\end{equation}
We start from the hyperfine structure of $^{133}Cs$
which is used as a frequency standard. Using eqs.(\ref{A},\ref{Cs}) 
we obtain 
\begin{equation}\label{VCs}
V(^{133}Cs)=\alpha^{2.83}(\frac{m_q}{\Lambda_{QCD}})^{0.11}
(\frac{m_e}{\Lambda_{QCD}})
\end{equation}
Here we have taken into account that the proton mass 
$m_p \propto \Lambda_{QCD}$.
For hyperfine transition frequencies in other atoms we obtain
\begin{equation}\label{VRb}
V(^{87}Rb)=\alpha^{2.34}(\frac{m_q}{\Lambda_{QCD}})^{-0.06}
(\frac{m_e}{\Lambda_{QCD}})
\end{equation}
\begin{equation}\label{VH}
V(^{1}H)=\alpha^{2}(\frac{m_q}{\Lambda_{QCD}})^{-0.09}
(\frac{m_e}{\Lambda_{QCD}})
\end{equation}
\begin{equation}\label{VD}
V(^{2}H)=\alpha^{2}(\frac{m_q}{\Lambda_{QCD}})^{-0.04}
(\frac{m_s}{\Lambda_{QCD}})^{-0.23}
(\frac{m_e}{\Lambda_{QCD}})
\end{equation}
\begin{equation}\label{Hg}
V(^{199}Hg^+)=\alpha^{4.3}
(\frac{m_s}{m_q})^{0.11}
(\frac{m_e}{\Lambda_{QCD}})
\end{equation}
\begin{equation}\label{Yb}
V(^{171}Yb^+)=\alpha^{3.5}
(\frac{m_s}{m_q})^{0.11}
(\frac{m_e}{\Lambda_{QCD}})
\end{equation}
\begin{equation}\label{Cd}
V(^{111}Cd^+)=\alpha^{2.6}
(\frac{m_s}{m_q})^{0.11}
(\frac{m_e}{\Lambda_{QCD}})
\end{equation}
  Now we can use these results to find limits on variation
of the fundamental constants from the measurements of
the time dependence of hyperfine structure intervals.
The dependence of  ratio of frequencies $A(^{133}$Cs)/$A(^{87}$Rb)
 can be presented in the following form
\begin{equation}\label{CsRb}
X(CsRb)=\frac{V(Cs)}{V(Rb)}=\alpha^{0.49} [m_q/\Lambda_{QCD}]^{0.17}
\end{equation}
Therefore, the result of the measurement \cite{marion}  may be presented
as a limit on variation of the parameter $X$:
\begin{equation}\label{limitCsRb}
 \frac{1}{X(CsRb)}\frac{dX(CsRb)}{dt}=
 (0.2 \pm 7) \times 10^{-16}/year
\end{equation}
Note that if the relation (\ref{mQCD}) is correct, variation of $X(CsRb)$
would be dominated by  variation of $[m_q/\Lambda_{QCD}]$.
 The relation (\ref{mQCD}) would give
$X(CsRb) \propto \alpha^{7}$.

For $A(^{133}$Cs)/$A$(H) we have
\begin{equation}\label{CsH}
X(CsH)=\frac{V(Cs)}{V(H)}=\alpha^{0.83} [m_q/\Lambda_{QCD}]^{0.2}
\end{equation}
Therefore, the result of the measurements \cite{Cs}  may be presented
as
\begin{equation}\label{limitCsH}
 |\frac{1}{X(CsH)}\frac{dX(CsH)}{dt}|<
 5.5 \times 10^{-14}/year
\end{equation}

For $A(^{199}$Hg)/$A$(H) we have
\begin{equation}\label{HgH}
X(HgH)=\frac{V(Hg)}{V(H)}\approx \alpha^{2.3}[m_s/\Lambda_{QCD}]^{0.1}
\end{equation}
The result of measurement \cite{prestage}  may be presented
as
\begin{equation}\label{limitHgH}
 |\frac{1}{X(HgH)}\frac{dX(HgH)}{dt}|<
 8 \times 10^{-14}/year
\end{equation}

 In Ref. \cite{Karschenboim} the limit on variation of the ratio
of hyperfine transition frequencies  $^{171}$Yb$^+$/$^{133}$Cs has
 been obtained (this limit is based on measurements \cite{Fisk}).
Using eqs. (\ref{VCs},\ref{Yb}) we can present the result as a limit
on $X(YbCs)=\alpha^{0.7}[m_s/\Lambda_{QCD}]^{0.11}$:
\begin{equation}\label{limitYbCs}
 \frac{1}{X(YbCs)}\frac{dX(YbCs)}{dt} \approx -1(2)
 \times 10^{-13}/year
\end{equation}

 The  optical clock transition energy $E(Hg)$ ($\lambda$=282 nm)
in Hg$^+$ ion  can be presented in the following form:
\begin{equation}\label{E}
E(Hg)=const \times [\frac{m_e e^4}{\hbar ^2}] F_{rel}(Z \alpha)
\end{equation}
 Numerical calculation
of  the relative variation
of $E(Hg)$ has given  \cite{dzuba1999}:
\begin{equation}\label{dE}
\frac{\delta E(Hg)}{E(Hg)}=-3.2\frac{\delta \alpha}{\alpha}
\end{equation}
This corresponds to $V(HgOpt)=\alpha^{-3.2}$. 
Variation of the ratio of the Cs hyperfine splitting $A(Cs)$
to this optical transition energy is described by
\begin{equation}\label{CsHgE}
X(Opt)=\frac{V(Cs)}{V(HgOpt)}=\alpha^{6}[m_q/\Lambda_{QCD}]^{0.11}
 [m_e/\Lambda_{QCD}]
\end{equation}
The work \cite{bize}
gives  the limit on variation of this parameter:
\begin{equation}\label{limitCsHgE}
 |\frac{1}{X(Opt)}\frac{dX(Opt)}{dt}|<
 7 \times 10^{-15}/year
\end{equation}

Molecular rotational transitions frequencies are proportional
to $m_e/m_p$, therefore one  should assume $V=m_e/\Lambda_{QCD}$.
For vibrational molecular transtions $V=(m_e/\Lambda_{QCD})^{1/2}$.
Comparison of Cs hyperfine standard with $SF_6$ molecular vibration
frequencies
was discussed in Ref. \cite{Chardonnet}. In this case
  $X(CsVibrations)= \alpha^{2.8}[m_e/\Lambda_{QCD}]^{0.5}
[m_q/\Lambda_{QCD}]^{0.1}$.    

The measurements of hyperfine constant ratio in different isotopes
of the same atom depends on the ratio of magnetic moments and 
is sensitive to $m_q/\Lambda_{QCD}$. For example,
it would be interesting to measure the rate of change for
 hydrogen/deuterium ratio where $X(HD)=[m_q/\Lambda_{QCD}]^{-0.05}
[m_s/\Lambda_{QCD}]^{0.23}$. 

 R. Walsworth suggested
 to measure the ratio of the Zeeman transition
 frequencies in noble gases which gives us time dependence of  ratio
of nuclear magnetic moments. Consider, for example $^{129}$Xe/$^3$He.
 For $^3$He the magnetic moment is very close
to that of neutron. For other noble gases nuclear magnetic
moment is also given by valence neutron, however, there are significant
many-body corrections. For $^{129}$Xe the valence neutron is
in $s_{1/2}$ state which corresponds to the single-particle value
 of nuclear magnetic moment $\mu=\mu _n= -1.913$. The measured value
 is $\mu= -0.778$.
The magnetic moment of the nucleus changes most efficiently
 due to the spin-spin interaction because valence neutron 
transfers a part of its spin $<s_z>$ to core protons and proton magnetic
moment is large and has opposite sign. In this approximation 
$\mu=(1-b)\mu_n + b \mu_p$. This gives b=0.24 and 
the ratio of magnetic moments $Y\equiv \mu (^{129}$Xe)/$\mu (^3$He)$ \approx
0.76 + 0.24 g_p/g_n$. Using eqs.(\ref{gps},\ref{gns}) we obtain
an estimate
for the variation of this ratio $\delta Y/Y \approx 0.1 \delta m_s/m_s$. 
 Therefore, in this case one can measure variation of
 $\mu (^{129}$Xe)/$\mu (^3$He) corresponding to variation of
 $X=[m_s/\Lambda_{QCD}]^{0.1}$.   

Note that an accuracy of the results presented in this paper
depends strongly on fundamental constant we are studying.
The accuracy for the dependence on $\alpha$ is few percent.
The accuracy for $m_q/\Lambda_{QCD}$ is about $30\%$, it is limited
by the accuracy of a single-particle approximation for nuclear
 magnetic moments. The results for $m_s/\Lambda_{QCD}$
should possibly be treated as  order of magnitude estimates.
However, the accuracy here may be improved  using proper
QCD calculation \cite{Thomas3}. Relation   
(\ref{mQCD}) between variation of $\alpha$ and  $m/\Lambda_{QCD}$
has been used as an illustration only.

The author is grateful to C. Chardonnet,
 S. Karshenboim, D.B. Leinweber, A.W. Thomas and
R. Walsworth for valuable discussions.
This work is  supported by the Australian Research
Council.

\end{document}